# The nature of representation in Feynman diagrams

Mauro Dorato[1]
Emanuele Rossanese[2]

Abstract

After a brief presentation of Feynman diagrams, we criticizise the idea that Feynman diagrams can be considered to be pictures or depictions of actual physical processes. We then show that the best interpretation of the role they play in quantum field theory and quantum electrodynamics is captured by Hughes' Denotation, Deduction and Interpretation theory of models (DDI), where "models" are to be interpreted as inferential, non-representational devices constructed in given social contexts by the community of physicists.

**1. Introduction**

The aim of the paper is to discuss the alleged representational role of Feynman diagrams (FDs) in the context of quantum field theory and in particular in quantum electrodynamics.

Physicists working in particle physics use FDs to calculate the outcomes of interactions between particles. However, it is still debated whether FDs are only a convenient tool to help the calculations or have instead a representational role of some kind. In the latter hypothesis, one could claim, for example, that they are pictorial representations of physical processes involving interactions between charged particles.[3] The problem with this claim is that in the context of quantum field theory and quantum electrodynamics, FDs seem to play only an algorithmic role. More precisely, while it is true that there is a sense in which FDs help to visualize what is going on during an interaction between charged particles, they do so only in a *heuristic way*. Our conclusion will therefore be that FDs are just important tools to simplify the calculations and we will defend this claim by adopting Hughes's theory of scientific *models* (Hughes 1997).

---

∗  We are indebted to the anonymous referee for help in clarifying our thoughts. We also gratefully thank David Atkinson for his encouragement and for having carefully read a previous version of this paper. We are also truly indebted to Letitia Meynell for her careful reading of a previous draft, which prevented some misunderstandings of her 2008 paper (Meynell 2008). The fact that we did not endorse all of her suggestions may be at our own peril.

1 Department of Philosophy, Communication and Media Studies, University of Rome 3, via Ostiense 234, 00146, Rome, Italy.
2 Department of Philosophy, Communication and Media Studies, University of Rome 3, via Ostiense 234, 00146, Rome, Italy.
3 For the notion of representation, see van Fraassen (2008).

The paper is structured as follows. In the second section, we will briefly illustrate the formalism of FDs and explain why certain of their features can lead to the mistaken belief that they can act as a *pictorial representation* of physical processes occurring in spacetime. We will also stress the aspect of the formalism that instead points toward an instrumentalistic account of the role of FDs. In the third section, we will briefly sketch the main problems encountered by the claim that FDs are pictures of the physical processes that they model. This criticism presupposes a general, albeit very sketchy account of what counts for a "pictorial representation". Clearly, an answer to the question "do FDs represent pictorially?" is going to depend on such an account. While taking for granted that pictures, like "games" in Wittgenstein's view (1953), cannot be exactly defined, we will suggest that *spatio-temporal intuitions* in the sense of Kant are the main characteristics of pictorial representations. In the fourth section, we will explain why, for the main purpose of the paper, framing the role of FDs in the context of a theory of scientific models originally proposed by Hughes (1997) seems a very plausible move. Finally we will mention in passing how this interpretation is reinforced by another account of models that is defended, for instance, by Suárez (2004) and van Fraassen (2008). In conclusion, the main claim of the paper is that the advantage of replacing the vaguer term "representation" with "model" does not consist in the fact that there is more philosophical agreement on what a model is (see Frigg and Hartmann 2016). Rather, it lies in the facts that (i) in the shared practice of physicists, models (FDs included) are regarded as algorithmic devises, thereby bolstering the account proposed here; (ii) the dispute between the friends and foes of the claim that FDs are pictures may be motivated by the merely contingent and psychological fact that physicists ─ among them the main creators of quantum electrodynamics (from now onward QED), Feynman and Dyson ─ rely on two different styles of scientific thinking.

## 2. Feynman Diagrams from A to B[4]

In order to pose the problem of the representational power of FDs, we need to sketch in a qualitative, brief but as-precise-as-possible way, the role that they have in quantum field theory. FDs were originally proposed by Richard Feynman (1949a, 1949b) to describe and calculate processes of interaction and scattering in the specific context of QED. Quantum electrodynamics is a quantum field theory that describes the fundamental interactions between photons and charged fermions in terms of scattering and bound-state processes. As is well known, QED describes the interactions between two quantized fields: the *electromagnetic field* (the Maxwell field) and the *electron-positron field* (the Dirac field).

In order to consider interactions in the context of quantum field theory, one has to formulate a perturbation theory by splitting the Lagrangian that describes the quantum field in two parts: the *free part*, that refers to non-interacting particles, and the *interaction part*, that refers to interactions. The formal tool that enables to calculate the interactions starting from an *initial free state* |IN> to a *final free state* |OUT>, is the S-matrix. It is important to note that both the initial and final free states are definable only in the *limit of time going to infinity*. That is, the |IN> and |OUT> states describe the physical world only when $t \to \pm \infty$. In evaluating the question that we are after, we ought not to forget the presence of this evident idealization. The calculation of the S-matrix expansion is often very difficult and the FDs have the fundamental role to make it simpler.

Accordingly, already Dyson (1948) considered the FDs as a set of diagrams with the relevant rules of calculations that have the aim to compute matrix elements for scattering processes. In the context of quantum electrodynamics, Dyson himself provided a formal derivation of the FDs for an initial (and final) state of a charged particle in a process in which there are no photons: each term in the matrix element can be associated to a *graph* such that there is a one to one correspondence between types of matrix elements and graphs. The combination of these two ingredients yields a FD.[5]

---

[4] The title of this section echoes that of Geroch's book *Relativity from A to B*, Chicago, 1978.
[5] In the secondary literature, this algorithmic role of the FDs has been stressed by Kaiser (2000, 61), who claims that FDs are nothing but "conventional representational schemes with no pretentions to picturing actual particles' real scattering".

## 3. Are Feynman Diagrams pictures?

In order to argue in favor of the claim that FDs are only a convenient formal tool to calculate the expansion of the S-matrix in virtue of their correspondence with matrix elements, we should first of all consider the arguments that may lead one to believe that FDs represent in a pictorial way. A general account of what counts for "pictorial representation" can be suggested by the way the typical FD is used, in particular if one considers that the FDs are usually *drawn* in a plane, where the horizontal axis represents space and the vertical axis represents time.[6] Moreover, FDs represent fermions with *straight lines* and bosons with *wavy lines* (FIG.1).

Additional, *prima facie* evidence for the claim that the FDs represent scattering processes pictorially is reinforced by the fact that (i) each fermionic line preserves its spatiotemporal orientation, which represents whether it is at the *origin* of the interaction or its *result* and that (ii) basic interactions are represented by a *vertex*, which is defined as the conjunction of at least three lines. These three lines stand for two fermionic particles and a bosonic particle, which represents the *minimal interaction* among the former. To each *vertex*, also suggesting a pictorial image, is associated a *coupling constant* that characterizes the interaction – for example, the electric charge in the context of QED. It is worth noting here that the idea of a coupling constant seems also to suggest the claim that in this context one is representing *a law of nature*, while the products of interactions might be regarded as *events* in spacetime. To conclude this brief description of the notion of vertex, it is important to remember that at each vertex the relativistic stress-energy tensor is conserved.

As a rough and ready general characterization of what counts for a pictorial representation, we will assume that the pictorial nature of any representational vehicle depends on the possibility to regard its target as a local matter of fact or as a local process that takes place *in spacetime*. For instance, a particular instance of this aspect of pictorial representations is the preservation of angles between the image of the vehicle and that of the target, as it occurs in the case of maps, photographs and "faithful" hand-made portraits. Also caricatures belong more or less to this class, where similarity can come in degrees.

Even if this criterion is too strict, we can use it as a first stab toward a more general characterization, which was implicit in the reference to space and time at the beginning of this paragraph. A pictorial representation in fact is grounded in the possibility of *visualizing* local processes or matters of facts thanks to what Kant called the pure a priori forms of our intuition of all possible phenomena, namely space and time. As we will see below, this element of visualizability is very relevant to explain the uncontroversial role played by the FDs in *learning* and in *understanding* the theory following the paradigmatic example of Minkowski's diagrams (Kaiser 2000). This pedagogical element is very plausibly grounded in our a priori "cognitive structure", as one could reformulate – in a partially faithful way – the above-mentioned principles of Kant's transcendental aesthetic. Such a structure is what led Schrödinger – in the wake of the neo-Kantian tradition of Helmholtz, Boltzmann and Hertz – to claim that what cannot be represented via a *Bild* in spacetime "cannot be understood at all" (quoted in Beller 1997, 427).

---

6 Some physics textbooks prefer the opposite convention, and represent time with the horizontal axis and space with the vertical axis.

Of course, after having granted that human beings are basically visual animals – since they extract most of the information on the external world through their eyes –, it would be too reductive to equate "understanding" a physical process with "visualizing it in space and time". However, even if the FDs could *not* be regarded as a pictorial representation of a spatiotemporal process in the rigorous sense of the word, one could still defend the weaker claim that they could act as *props* to try to understand what happens in the physical world (Meynell 2008).

Since we will discuss this claim in a later section, here it is appropriate to note there are at least two arguments against the claim that FDs can be regarded as pictures of the physical world in a strict sense. The first has been already advanced by Brown, who noted that as a consequence of the Heisenberg uncertainty principle, particles lack a well-defined spatiotemporal trajectory (Brown 1996). A difficulty with this criticism, however, is that it is interpretation-dependent, given that in Bohmian mechanics, for instance, particles always possess a definite position and velocity.

A second, more technical difficulty against the view that FDs can be considered as pictorial representations of real physical processes in spacetime depends on some specific, practical aspects of the physicist's use of FDs.

In fact, if we assign the |IN> state different particles, the *same* fundamental diagram can represent different physical processes, as it is possible to see in (FIG. 1). It is true that the two FDs have a different spatiotemporal orientation on the plane, but here the point is that the two FDs are *structurally equivalent* (in the sense that they are both composed by four straight lines connected by a wavy line, that is, both are composed by two vertexes). This means that the only difference between them is a difference in spatiotemporal orientation, since one can be interpreted as the result of the rotation of the

other. However, physics tells us that these two pictures are *structurally the same* in the essential sense that are invariant for rotations. Consider a map: if we rotate it of 90 degrees, the internal relations between the elements of the map remain the same. Yet, a particular FD regarded as a pictorial representation of a given physical process should differ from a FD representing another physical process.

As an example, consider the first FD on the left, which represents a process of diffusion of electrons and protons, while the second FD on the right represents the annihilation of an electron and a positron with the creation of a pair of a fermion and an antifermion. The term $\sqrt{\alpha}$ in the diagram is the coupling constant, the $\gamma$ is the *propagator* (identifiable with the exchange of a *virtual particle*), $1/q^2$ is the impulse transported by the photon, where $q$ is the charge of the particle. In other words, this case shows how regarding FDs as representations of physical processes in spacetime would imply that different physical processes have the same representation. As we shall see later in this section, this is just one of the problems that a pictorial interpretation of FDs has to face.

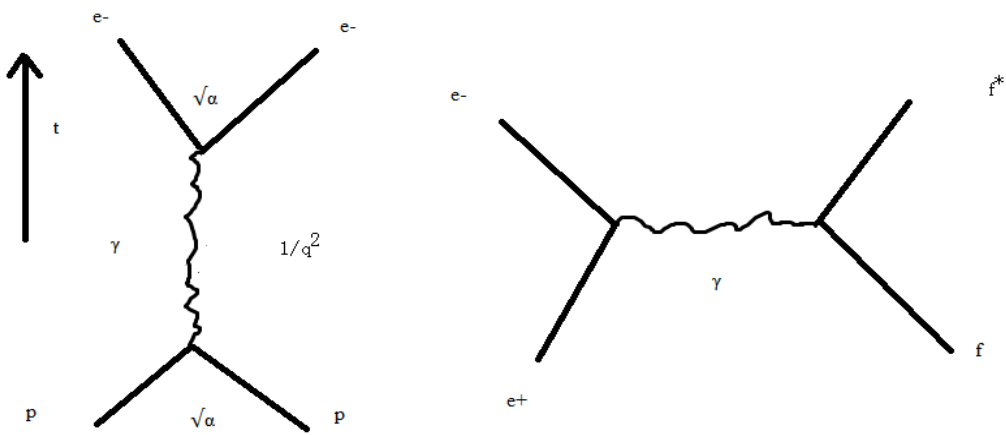

(FIG. 1)

As a second argument, in order to appreciate the fact that the FDs are merely bookkeeping devices, we should stress their role in calculating scattering processes within the so-called Feynman's *spacetime approach* to quantum electrodynamics. If we consider an electron-electron scattering process, we will have a virtual photon propagating *all over Minkowski spacetime*, i.e. tracing all the possible paths that connect the first electron to the other electron. This propagation is represented by a time-ordered product in the S-matrix. This virtual photon (Feynman's propagator) is assumed to be *created* at a spacetime point $x_1$ and to be *annihilated* at a spacetime point $x_2$. To calculate the S-matrix, one then needs to consider the contribution of all possible localized interactions of the Dirac fermionic and Maxwell bosonic fields, in the way in which they are connected by the virtual photon propagator. Technically speaking, one has to integrate over the probability amplitudes of all these possible paths.

Moreover, it must be noted that the interaction that we are now describing results from the contribution of the propagation of the virtual photon from one electron and the other, *and viceversa.* Since FDs can be seen as a representation of a global spatiotemporal description of scattering processes all over Minkowski spacetime, how can this feature be understood as a picture of a real, localized physical process? In other words, "overall" means that we do not have a description of the interaction taking in a local region of spacetime, that is, we do not have an "in space" and "in (through) time" description of the interactions. A process, exactly as a particle, must be contained in a *local* region of spacetime!

As a consequence, it is much more plausible to assume that a FD is just a "black box" that helps to calculate the complex function from an |IN> to an |OUT> state associated to a certain interaction.

Given the conclusion of this section, we have to explain their allegedly "pictorial" role in the physics textbooks. This introduces the topic of the next section, which involves the heuristic role of the FDs. Kaiser (2000) for one regards FDs as *conventions,* but not chosen at random, since he argues that the way in which they are introduced and explained to physics students relies on a *pedagogically useful* comparison with Minkowski diagrams.[7] However, while a painting need not be a representation *of* something else, namely a picture of something, the case of FDs is fundamentally different: what matters in our case is whether they are pictures *of* something, not that they can be used, either pedagogically or heuristically, as props for our imagination, or make-believe (see below).

---

[7] In what follows, we will not venture into evaluating Kaiser's interesting claim that the wrongheaded association of *realism* to FDs in 1950s and 1960s has been suggested by their similarity with the "real" photographs of "real" particles in the bubble chambers.

## 3. In what sense could FDs be taken to represent? Feynman diagrams as props of our imagination

Even if we reject the claim that FDs are pictorial representations of physical process in the rigorous sense of "picture" specified above, the main problem to face when we ask the question whether and in what sense could FDs represent is our lack of understanding of a more general question, namely: What does it mean to claim that a physical theory represent or fails to represent the world? This difficulty has already been stressed by Walton, when he refers to "the confused profusion of senses and nonsenses" (Walton 1990, 3) in which the term "representation" is used. Depending on various philosophical positions on the nature of scientific representations, one gets different answers. And typically, even if we had a widely shared explication of the notion of scientific representation (which we don't), the application of this *general* explication to two *different* physical theories may end up leaving us with different answers.

In our case, not only is the issue of the representational power of FDs subject to the vagueness of the concept of "scientific representation", but the problem is made even more complicated by the simple remark that FDs *prima facie* are *not* a physical theory of phenomena, in the sense in which, for instance QED is. If they were, one could try to claim that FDs are families of models of the quantum world that represent in some non-pictorial way the relevant phenomena, in the same sense in which physical theories in general represent.

Neither can be of help the hypothesis that FDs are *laws* governing the interaction between electrons and photons or, more generally, other kinds of interactions, wherever this term applies. On the one hand, as hinted above, scatterings are certainly physical processes taking place in spacetime and coupling constants derive from physical theories. On the other hand, however, the issue whether the FDs are bookkeeping devices or something more cannot be resolved by trying to understand the nature of natural laws, since it is always possible to defend a view of laws that regards them as mere inference tickets.

Luckily here we can avoid these difficulties by focusing on the more circumscribed question whether the FDs represent in a sense that is weaker than the sense of "pictorial representation" already criticized in the preceding section, namely on Meynell's interesting claim (2008) that FDs act or can act as props for the scientist's imagination. This is the more general sense of representation that will be referred to in what follows. However, in order both to defend our instrumentalist take on FDs and reinforce Brown's view on the matter (1996), we need to specify in a clearer way the conditions under which it would be legitimate to claim, along with this author, that scientific representations in general (and FDs in particular) could act as a prop for our imagination. Let us refer to this weaker sense of picture as "quasi-picture". Later we will defend an inferentialist view of the nature of scientific models, in which any representational role is avoided. It must be granted that there is a continuous transition between clear cases of pictorial representations and cases in which mere inkblots on a flat surface have no pictorial character whatsoever. Rothko's paintings, for instance, belong to this latter case, but it seems to us that they would still count, on Meynell's criterion, as a prop for our imagination or quasi-pictures. Of course, we could *see* Rothko's painting *as* …, but whatever 'we see as' risks being subjective or not intersubjectively shared. We all see the same profile of inkblots on a flat piece of paper, but what matters in Rorschach's tests is what one *see as*, and this is why they are used as a test for personality traits. Meynell of course knows that if FDs were like Rorschach's spots, we could not treat them as quasi-pictures of scientific representations, because they must prompt the *same* quasi-picture or image to everybody. On the other hand, there are genuine pictures of non-existent objects (Santa Claus)[8], and possibly *all* scientific models refer to fictional entities. However, this point would just reinforce the claim that FD do not pictorially represent local actual physical processes or matters of fact in spacetime. This is the conclusion that we are after.

The charge of subjectivity of the representation force of "prop for the imagination" yielded by "quasi-pictures" – is strengthened by the fact that there is an important subjective element in the psychology of scientific imagination, given by the existence of different *styles of scientific thinking* This fact is another piece of evidence in favor of an instrumentalistic view of FDs. Interviewed by the mathematician Jacques Hadamard, some mathematicians and physicists reported that they "rarely think in words at all", but rather take advantage of visual and other kinds of mental images, often pictures.[9] Others, on the contrary, seem to rely on purely symbolic or linguistic ways of thinking. Interestingly, given what we know about his style of doing science, Dyson belong to the latter type of scientists while

---

8  This objection has been raised to us by Meynell in private correspondence.
9  Einstein for once was aware of this difference, when in a letter to Hadamard he wrote "[t]he words of the language, as they are written or spoken, do not seem to play any role in my mechanism of thought. The psychical entities which seem to serve as elements in thought are certain signs and more or less clear images which can be 'voluntarily' reproduced and combined. The above mentioned elements are, in my case of *visual* and some of a muscular type. Conventional words or other signs [presumably mathematical ones] have to be sought for laboriously only in a secondary stage, when the associative play already referred to is sufficiently established and can be reproduced at will." (Hadamard 1945, 142-143, italics added).

Feynman to the former, something that could explain why they held different ideas about how the diagrams should interpreted and used. "From the very beginning, Feynman and Dyson held different ideas about how the diagrams should be drawn, interpreted, and used. For Feynman, doodling simple spacetime pictures preceded any attempts to derive or justify his new calculational scheme. To Dyson, the diagrams could be of any help only if they were first derived rigorously from a specific field-theoretic basis. To Feynman, his new diagrams provided pictures of actual physical processes, and hence added an intuitive dimension beyond furnishing a simple mnemonic calculational device. To Dyson, the drawings were never more than 'graphs on paper', handy for manipulating long expression." (Kaiser 2005, 175–176). [10]

Of course these biographical facts have no weight in evaluating the *normative* question we are trying to establish, but raise in any case three points that are rather important for the main claim of this paper.

1. The first is that part of the *philosophical* discussion about whether FDs pictorially represent or not has its roots in the two main protagonists' different imaginative or cognitive styles. However, the *fact* that some physicists (Feynman included) consider FDs as pictures or props for his imagination while other (Dyson included) do not prevents one from claiming that FDs represent as pictures, since whether something represents pictorially or not cannot just be a matter of different cognitive styles of reasoning. The other two points are more linked to our instrumentalistic stance.

2. Pictorial or quasi-pictorial props for our imagination may play a role in inventing a physical theory or a new calculation tool (*in the context of discovery*), but since their justification and effectiveness lies in something else (theorems for instance, or precise predictions expressed in symbols within the context of justification), these images can only be considered as instrumental to suggesting the theory. As Kaiser has insisted, the act of baptism that led to the discovery of a certain technique in the context of discovery is not forgotten by later scientists, but becomes in some sense a consolidated way to present the theory even in the context of justification, even though this is compatible with the fact that the role of FDs is purely pedagogical.

3. It seems to us that Meynell's position does not completely avoid the risk of endorsing a purely subjective view of imagination, despite the fact that principles of generation of new hypotheses may possess some degree of intersubjectivity, given by the heuristic strategies that are typically used by scientists sharing the same paradigm. However, here we are not making general claims about normal science, but are rather discussing the specific case of FDs.

---

10 Feynman (1997), as he himself tells us in his *Surely you are joking Mr. Feynman?*, thought in pictorial images when, for example, he tried to prove visually a theorem that his colleagues mathematicians had proved in symbolic sequences.

Meynell could reply to our charge of subjectivity by drawing a distinction between "pictures" and "depictions" of phenomena. Consider the following quotation: "depiction differs [from pictures] in that it is through seeing the picture that we imagine ourselves seeing the pictured state of affairs" (Meynell 2008, 50). This means that when we look at a FD, we imagine ourselves seeing the quantum phenomenon we are trying to describe, much in the sense described above by Einstein (see note 9) and Feynman. But if our images are off target, FDs do not depict at all a real physical process.

If we need a criterion of correctness for our imagination in order to claim that FDs represent as depictions or picture, we must disagree with the following claim: "the question of how and whether particular pictures denote is certainly an important question, but it is a different question to the question about how and whether marks on a surface – say, Feynman diagrams – pictorially represent." It is a different question, but the most important question, it seems to us, is the former (i.e. a normative question that Meynell does not raise) and not the latter: if by looking at a FD we imagine ourselves seeing a true quantum phenomenon, how can we be sure that we are seeing something out there and not merely hallucinating? Imagination seems too weak a criterion for a pictorial representation in science, so that we are pushed toward a normative criterion that tells us that we are indeed depicting something and note merely imagining something that might not exist by relying on a picture that is a prop for our imagination. In a word, it is easy to claim that FDs are pictorial representations of physical processes if denotation of some sort is not taken to be necessary for a "depiction".[11]

Finally a reason to be skeptical about an objective pictorial component of FDs is given by the plausibility of the thesis that pictorial representations, and scientific representations in general, have an intentional component and a conventional element, that draws out attention to the context of use (Callender and Cohen 2006). "I use a picture M with the intention to represent P" is analogous to "I use the black ball that is about to collide with the red ball to represent a scattering of particles", or even, "I use a salt shaker and a plate to represent the Moon orbiting the Earth". This intentional component may well force one to conclude that whether or not FDs are pictures of something or not may depend on the particular use or aim at hand and therefore by the intention of the speaker.

In other words, Meynell's brilliant attempt to separate the question of representation from that of denotation faces some objections. In the case of a physical theory a pictorial or quasi-*pictorial* representation must be a depiction of something beyond itself. That FDs can be props for a visual imagery is certainly important but we hope to have shown that this is at most a purely de facto, subjective and psychological phenomenon. As such, the amazing strength of FDs as calculational device capable of predicting with the utmost precision the magnetic moment of the electron, cannot depend on the fact that they are literally pictures of scattering processes or props for our imagination. In other words, a representational account of the FDs – in the sense of "representation" discussed above – cannot be regarded as the best explanation of QED's predictive success.

---

[11] Meynell's non-factive view of representation seems appropriate for art, but not for science.

**4. A model-theoretical, inferentialistic view of representation of Feynman diagrams\***

If the notion of pictorial and quasi-pictorial representation cannot be a valid interpretation of the role of FDs in the context of quantum field theory and quantum electrodynamics, we would like to suggest a different, anti-representational perspective. Such a perspective which supports an instrumentalist view of FDs, relies on a particular understanding of the concept of *scientific model*, the *inferentialistic* one, that seems well-suited to account for the actual *use* of the diagrams on the part of physicists.[12]

While we are well aware that, in general, a sociological fact is not sufficient to ground a normative, epistemological hypothesis, in our case a failure to take into account the concrete practice of physicists in their employment of the FDs is bound to fail. Therefore, our claim is not that we understand "model" better than "representation": the question on the nature of scientific models cannot be solved in the present context (see Frigg and Hartmann 2016 for an informed survey) and there is no universally shared position on the nature of scientific models.

Our shift of attention on scientific models, however, is justified by the remark that one of the most interesting but neglected questions surrounding the FDs is, as suggested above, why they are so effective in predicting physical phenomena. It is well known that the prediction of the magnetic moment of the electron, if compared with the experimental data, is possibly the most precise every obtained in the history of physics. If this is the problem on which future research on FDs should concentrate, we suggest that the relationship between the mathematical model and the data model might prove crucial. The problem whether FDs can be regarded as mathematical models of experimental data would pose the issue of their relationship with the physical world under a different light. In this perspective, the main questions raised above would be translated into the following one: if the FDs were treated as mathematical models, in what sense could they denote the physical world?However, the predictive, algorithmic component of the FDs, as it results from the antirepresentationlist conclusion defended above as well as from the concrete practice of the physicists, seems to calls for some sort of *denotation, deduction and interpretation* model of the phenomena, of the kind defended by Hughes (DDI's model). Denotation might be interpreted as the data models corresponding to |IN> state, the deduction part corresponds the algorithm used to calculate the element of the S-matrix via the help of FDs, and the FDs' |OUT> state is Hughes' interpretation stage.

---

12 Mattuck (1967) claimed that FDs have a *quasi-physical* nature, since they do not really represent something physical, but neither are they simple mathematical devices. This claim, which here cannot be evaluated, relies on a piecemeal approach to the question of scientific realism (Fine 1991), that partially inspires also this paper.

The following example is a simple but good illustration of our claim, and is taken from Feynman's celebrated, popular book on QED, where he explains the conceptual roots of his approach to quantum field theory, essentially involving a sum over all amplitudes (1985, 26) (FIG. 2)

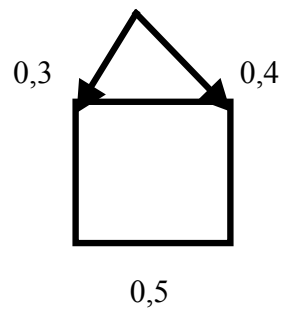

(FIG. 2)

FIG. 2 is the mathematical model that is employed in our example. Let the arrows denoted by 0.3 and 0.4 be the probabilities of a particular process (their amplitude). This corresponds to the process that Hughes calls *denotation*, the first step of his three-fold theory, which in our view implies *embedding a property of the physical system in a mathematical, data model*.

The second step is purely inferential and deductive, since it consists in calculating the length of the diagonal and then squaring the number. This is the process that Hughes calls *deduction*, which is *epistemically* crucial in that it allows us to reason *via* the vehicle by using it as a "surrogate" for the actual physical systems (see also Suárez 2004). The mathematics used in this second step increases our information about the original system (Azzouni 2004) but, as claimed above, is devoid of realistic components. In Hughes words: "from the behavior of the model we can draw hypothetical conclusions about the world over and above the data we started with" (Hughes 1997, S331).

The final step, which in a sense is the converse of the first, is *interpreting* the result (0,5), obtained via the algorithmic prescription, as the probability of the event we are interested in: this step transforms the information provided by the algorithm into something that can be subject to confirmation or disconfirmation.

If we summarize the main points featuring in Hughes' view of models as they are applied to our case via the example above, we easily realize that the link between the first and the third step listed below is purely inferential:

1. Denote the |IN> or the probability of particular physical processes by the arrows (the mathematize data model ( → *Denotation by a mathematical model* );
2. Make the relevant calculations by summing the arrows in the appropriate way ( → *Deduction via FDs* );
3. Interpret the square of the diagonal of the arrows as the probability of the data/event we are interested in ( → *Interpretation of the end product of the calculation*, namely the |OUT> ).

Two important remarks are appropriate at this point. First, the conventional elements in the DDI model play an important role, since the choice of denoting a certain physical property by a certain component of a mathematical model (embedding via arrows, for example) is dictated by reasons of convenience and tractability of the calculations. Second, the DDI model functions in many scientific applications and, in our specific case study, it also is very appropriate to describe the role played by FDs in the physicists' practice.

One might want to put forward the point that the assumption that scattering processes are somehow correctly *denoted* – in Hughes' sense – by the relevant FDs is the best explanation for their predictive success. This might be taken to imply that such diagrams model or denote, in some non-pictorial but robust, structuralist way (in the sense of Da Costa and French's 2003),[13] concrete physical processes. In order to cast some doubts on realistic arguments of this sort, first of all we should notice that the FDs are not necessary to calculate the scattering of particles: *without* presupposing them, one can still calculate the scattering of the two particles *via* the integral of the particles; so in this case any sort of indispensability argument. Consequently, the empirical success of FDs cannot be explained by an inference to the best explanation involving their structuralist or similarity-based (Giere 1988) features. In fact, the diagrams, as other bits of applied mathematics, in our case have been devised with the specific aim of calculating more exactly the magnetic moment of the electron and in general the interactions between matter and light: no wonder that they are so successful! Inferences appealing to the fact that many mathematical models of theoretical physics are successful in order to argue that all successful models (Feynman's included) "depict" reality in a unique way – so that a uniquely describable representation relation stands out – are not convincing, because the algorithmic power of FDs may be completely independent of the real processes whose end result they enable us to calculate.

---

13 These authors do not use FDs to illustrate their structuralist view of mathematical modelling of pysical phenomena.

In order to add some pragmatic component to our claim, we conclude by mentioning Suárez's inferentialist account of models (2004), which is not specifically referred to FDs. His position, which is similar to Hughes', focuses more specifically on the community of speakers that use a model in order (i) to grasp the physical process that they are describing and (ii) communicate their analysis in the clearest possible way. According to Suárez, a model is in fact mainly an instrument that lets cognitively informed agents draw inferences about the target, but its denoting and interpretive component is highly contextual.

If sum, the best interpretation of the role of FDs in quantum field theory and quantum electrodynamics is captured by an account of their use *via* Hughes' DDI theory of models, where "models" are interpreted as inferential, non-representational devices constructed in given contexts by the particle physics community.